\documentclass[12pt,a4,reqno]{amsart}
\begin{document}

\title{The increase of Binding Energy and Enhanced Binding in Non-Relativistic
QED}

\author[Chen]{Thomas Chen$^1$}
\address{$^1$ Courant Institute  of Mathematical Sciences, New York University,
251 Mercer Street, NY 10012-1185, USA}
\email{chenthom@cims.nyu.edu}

\author[Vougalter]{Vitali Vougalter$^2$}
\address{$^1$ Department of Mathematics and Statistics, McMaster University,
1280 Main Street West , Hamilton, Ontario L8S 4K1, Canada}
\email{vougav@math.mcmaster.ca}

\author[Vugalter]{Semjon A. Vugalter$^3$}
\address{$^2$ Mathematisches Institut, LMU M\"unchen,
Theresienstrasse 39, 80333 Munich, Germany}
\email{wugalter@mathematik.uni-muenchen.de}
\setlength{\parindent}{0pt}
\date{}
\thanks {$^1$  Courant Institute  of Mathematical Sciences, New York University. 
Supported by a Courant Instructorship.}
\thanks {$^2$  Department of Mathematics and Statistics, McMaster University}
\thanks {$^3$ Mathematisches Institut, LMU M\"unchen}

\begin{abstract}
We consider a Pauli-Fierz Hamiltonian for a particle coupled to a
photon field. We discuss the effects of the increase of the binding energy and
enhanced binding through coupling to a photon field, and prove that both effects
are the results of the existence of the ground state of the self-energy operator
with total momentum $P = 0$.
\end{abstract}
\pagestyle{empty}
\setlength{\parindent}{0pt}
\setlength{\parindent}{1.5em}
\renewcommand{\baselinestretch}{1.0}
\newtheorem{thm}{Theorem}
\newtheorem{cor}{Corollary}
\newtheorem{lem}{Lemma}
\newtheorem{con}{Condition}
\renewcommand{\thecon}{\arabic{con}}
\newtheorem{rem}{Remark}
\newtheorem{dfn}{Definition}

\renewcommand{\theequation}{\thesection.\arabic{equation}}
\maketitle

\bigskip
\section { Introduction. }
We consider a charged particle coupled to a photon field that interacts with an
external potential in nonrelativistic QED.
This system will be described by a
Pauli-Fierz Hamiltonian, whereas 
neglecting the radiation effects, one obtains a corresponding 
Schr\"odinger operator. In the present paper, we discuss two closely related
questions:
\begin{enumerate}
\item
Does the interaction with a quantized radiation field increase binding abilities
of a potential (whether the Pauli-Fierz Hamiltonian can have a ground state if
the Schr\"odinger operator with the same potential does not)?
\item
If the corresponding Schr\"odinger operator has discrete spectrum, should the
binding energy (the difference between the infimum of the energy with and
without potential, measured in units $mc^2$, where $m$ is the bare electron mass) 
increase if the interaction with the radiation field is 
considered?
\end{enumerate}
We emphasize here that the asserted increase of binding energy holds with respect
to the bare electron mass. 
In physical experiments, the binding energy is usually measured in units 
$m_{phys}c^2$, where $m_{phys}$ is the rest mass of the free infraparticle 
(comprising the free electron together
with a cloud of low-energetic photons that it binds). 
In these units, the binding energy decreases.

Physical intuition suggests that the answer to both (1) and (2) should be
in the affirmative. The free infraparticle binds a larger quantity of 
low-energetic photons than the confined particle, and thus possesses a 
larger effective mass. In order for the particle to leave the potential well, 
an additional energetic effort, proportional to the difference of 
the two effective masses, is therefore necessitated, relative to 
the situation without coupling to the quantized electromagnetic field.

Recently, problems 1 and 2 were actively studied in the mathematical literature.
First, let us mention the paper by M. ~Griesemer, E. ~Lieb, M. ~Loss
~(\cite{GLL}), where the authors proved that the binding energy cannot be
decreased by the photon field. Another (and more important) achievement of
~\cite{GLL} is a criterion for a Pauli-Fierz Hamiltonian to have a ground state.
This criterion will be also used in the present paper. The investigation of
enhanced binding (Problem 1) was started by F. ~Hiroshima and H. ~Spohn
(~\cite{HS}), who considered the Pauli-Fierz Hamiltonian in the dipole
approximation. In this  approximation, the dependence of the magnetic vector
potential on the coordinates of the particle is neglected. They proved the
existence of  enhanced binding for sufficiently large values of the coupling
parameter $\alpha$ (which is the fine structure constant, $\alpha \approx
\frac1{137}$ in nature) .

A different approach was implemented in~\cite{HVV}. The Pauli-Fierz operator without spin term ($ \sqrt{\alpha}\sigma \cdot B$) was studied with a potential, for which the corresponding Schr\"odinger operator does not have discrete eigenvalues, but which is very close to the appearance of the first eigenvalue.  The first step was to estimate the self-energy for small $\alpha $ with an error of the order $o(\alpha ^2)$, and then it was proved that by adding the potential, one gets a shift $C\alpha ^2$ of the infimum of the spectrum, which for small $\alpha$ implies the existence of the ground state~\cite{GLL}. This approach was further developed in a recent preprint~\cite{CH}, where the case of a particle with spin was considered.  

The increase of the binding energy for the Coulomb potential $e|x|^{-1}$ in a model situation, where the electron charge $e$ is constant, but $\alpha$ tends to zero, was proved by C.Hainzl~\cite{H}, by controlling the expansion of the ground state energy to order $o(\alpha ^2)$. To establish the corresponding result for the physical case $e = -\sqrt{\alpha} $ by the methods of~\cite{H}, one would have to control the expansion at least up to order $O(\alpha ^3)$. On the other hand, simple physical arguments show that this effect is caused by the form of the self-energy operator and does not depend on the coefficient of the potential. For fixed $\alpha$, the increase of the binding energy should exist for all values of $e<0$. 

It is nessesary to emphasize that the methods of~\cite{HVV} and~\cite{H} are asymptotic in $\alpha$, and that they can hardly be generalized in a manner to cover the physical case, where $\alpha$ is a fixed constant. Studying this case requires a different strategy, which is not  based on asymptotic expansions in $\alpha$. The work at hand is the first attempt to develop such methods. We prove two very simple theorems, showing that both effects take place if the self-energy operator, restricted to the states with total momentum $P=0$ (operator $T_0$), has a ground state. The proof of these two statements is based on direct variational estimates of the binding energy, and the results are independent of $\alpha$. Establishing the connection between the existence of the ground state of the operator $T_0$, and the existence of enhanced and increased binding is the main achievement of the present paper. 

The existence of the ground state of $T_0$ is a problem important for different aspects of nonrelativistic QED, and has been solved in~\cite{Ch}  for small $\alpha$. Applying the results of~\cite{Ch} and some generalizations thereof stated in the Appendix of the present paper, we immediately obtain a new, very simple proof of enhanced binding for small $\alpha$, in both the spin and spinless cases, as well as the proof of the increase of the binding energy for all potentials, for which the corresponding Schr\"odinger operator has a ground state. In particular, we prove the increase of  binding energy for the Coulomb potential ($e|x|^{-1}$), for all $e<0$.

\section { Definitions and main theorems. }
\setcounter{equation}{0}
The Hamiltonian  for an electron interacting with the quantized radiation field
and a given external potential $V(x)$, $x \in {\Bbb R}^3$ is
\begin{equation}
H = T + V(x) \ , \label{f1.1}
\end{equation}
where
\begin{equation}
 T= (p + \sqrt {\alpha} A(x))^2 + g\sqrt {\alpha}\sigma \cdot B(x) + H_f\ .
\end{equation}
We fix units such that $\hbar=c=1$ and the electron mass $m= \frac1{2}$, $\alpha =
e^2$ is the ``fine structure'' constant, $e$  is the electron charge. The
natural value of $\alpha$ is $\simeq \frac1{137}$, however, as usual ~\cite{BFS,
HS} we will think about $\alpha$ as a parameter in the operator $T$. The main
results of this paper  (Theorems ~1, 2) are true for all $\alpha >0$. An
artificial parameter $g$, which can attain the values of either $0$ or $1$ is
introduced to describe both the spin ( $g=1$ ) and the spinless  ( $g=0$ )
cases. As usual $\sigma = (\sigma _1, \ \sigma _2, \sigma _3)$ is the vector of
Pauli matrices, $p= -i\nabla _x$, $B(x) = $curl$A(x)$. The magnetic vector
potential $A(x)$ is given by
\begin{equation}
A(x) = \sum\limits_{\lambda = 1, 2}\int\nolimits_{{\Bbb R}^3}\frac{\chi (|k
|)}{2\pi |k |^{1/2}}\varepsilon _{\lambda}[a_\lambda (k)e^{ikx}+ a_{\lambda}^*
(k)e^{-ikx}]dk\ , \label{f2.1}
\end{equation}
where the operators $a_\lambda$, $a_\lambda ^*$ satisfy the usual commutation
relations
$$
[a_\nu (k),\ a_\lambda ^*(q)] = \delta (k-q)\delta _{\lambda, \nu }\ , \qquad
[a_\lambda (k),\ a_\nu (q)]=0\ .
$$
The vectors $\varepsilon _\lambda (k) \in {\Bbb R}^3$ are the two possible
orthonormal polarization vectors perpendicular to ~$k$.

The function $\chi (|k|)$ in ~(\ref{f2.1}) describes the ultraviolet cutoff on
the wavenumbers $|k|$. The only restriction on $\chi (|k|)$, which we need at
the moment,  is $\chi (|k|)\equiv 0$ for $|k|>\Lambda$ with some $\Lambda >0$.

The photon field energy $H_f$ is given by
$$
H_f = \sum\limits_{\lambda = 1,2}\int\nolimits_{{\Bbb R}^3}|k|a_\lambda
^*(k)a_\lambda (k)dk\ .
$$
Regarding the potential $V(x)$ we assume that $V(x)=V(|x|)$, $V(x)\in {\mathcal
L}_{2, loc} ({\Bbb R}^3)$, $|V(x)|\le C$ for $|x|\ge a$ with some constants
$C>0$, $a>0$ and $|V(x)|\to 0$ as $|x|\to \infty$. For $g=1$ the operators $T$
and $H$ are considered on the space
$$
{\mathcal H} = {\mathcal L}_2 ({\Bbb R}^3 ;\ {\Bbb C}^2)\otimes {\mathcal F}\ ,
$$
where $\mathcal F$ is the Fock space for the photon field.

If $g=0$ the corresponding space is
$$
{\mathcal H} = {\mathcal L}_2 ({\Bbb R}^3)\otimes {\mathcal F}
$$
According to ~\cite{Hi} the operator $H$ is semibounded from below and
essentially selfadjoint.

For an arbitrary  selfadjoint operator $\mathcal A$ let $s({\mathcal A})$ and
$s_{disc}({\mathcal A})$ be the spectrum and the discrete spectrum of $ü\mathcal 
A$
and
let
$$
E_0 = \inf s(T)\ , \qquad E_1 = \inf s(H)
\quad {\rm and }\quad \Delta _{E} = E_0  - E_1\ .
$$
To compare the binding energy in the presence of the photon field $(\Delta _E)$ 
and
the binding energy without photon field let us consider the Schr\"odinger
operator
\begin{equation}
h = - \Delta + V(|x|) \label{f3.1}
\end{equation}
with the same potential $V(|x|)$ as in (\ref{f1.1}). Denote by $-e_0$ the lowest
eigenvalue of the operator $h$ (if $s_{disc}(h)=\emptyset$, $e_0 =0).$ Obviously
$e_0$ is the binding energy for the Schr\"odinger operator. According
to~\cite{GLL}
\begin{equation}
\Delta _E \ge e_0§ . \label{f3.2}
\end{equation}
In the present paper it will be proved, that under some conditions, the strong
inequality
\begin{equation}
\Delta _E > e_0§      \label{f4.1}
\end{equation}
holds .

Suppose now that the potential $V(|x|)$ is short-range , $|V(|x|)|\le
c(1+|x|)^{-2-\delta}$, $\delta >0$ and its negative part is nontrivial. We can
write $V(|x|)$ as $\beta V_0(|x|)$ , where $V_0(|x|)$ is a short-range
potential, which we will keep fixed and $\beta >0$ is a coupling constant.
Denote by $\beta _0$  the minimal value of the coupling constant $\beta$, such
that for $\beta > \beta _0$, the operator $h$ with the potential $\beta 
V_0(|x|)$ has
nonempty discrete spectrum and by $\beta _1$ the minimal value of the coupling
constant such that for all $\beta > \beta _1$ operator $H$ with the same
potential has a ground state.

It will be proved that under the same conditions as we need for ~(\ref{f4.1})
$$
\beta _1 < \beta _0\ .
$$
This means that the photon field increases the binding abilities of
potentials. To formulate those conditions, we need to introduce some more
definitions.

First, notice that the operator $T$ is translationally invariant. It commutes
with the operator of the total momentum
\begin{equation}
P_{tot} = p_{el}\otimes I_f  + I_{el}\otimes P_f \ , \label{f.5.1}
\end{equation}
where $p_{el}$ and $P_f = \sum_{\lambda = 1,2}\int d^3k \,
ka_{\lambda}^*(k)a_\lambda (k)$ denote the electron and the photon momentum
operators respectively.

The Hilbert space $\mathcal H$ can be written as a direct integral
\begin{equation}
{\mathcal H} = \int\nolimits^{\oplus} d^3P {\mathcal H}_P\ , \label{f.5.2}
\end{equation}
where ${\mathcal H}_P$ are the fibre Hilbert spaces associated to the fixed 
values
$P$ of the conserved momentum, which are invariant under space and time
translations. For any fixed value $P$ of the total momentum the restriction of
$T$ to the fibre space ${\mathcal H}_P$ is given by the operator
\begin{equation}
T_P = (P - P_f + \sqrt{\alpha}A(0))^2 + \sqrt{\alpha}g\sigma \cdot  B(0) + H_f \ 
.
\label{f.5.3}
\end{equation}
The operator $T_P$ with $P=0$ (we will call it $T_0$) plays an especially
important role. 

The main theorems will be proved under the following assumption.
\begin{con}
There is an element $\psi _0 \in {\mathcal H}_0$, satisfying
\begin{equation}
T_0\psi _0 = E_0\psi _0 \; . \label{f.6.1}
\end{equation}
\end{con}
We note that Condition 1 contains two parts. For it to be satisfied, we
must first have 
$$
E_0={\rm inf}  s(T)= {\rm inf} s(T_0) \; ,
$$ 
and secondly, $E_0$ has to be in the point spectrum of 
the operator $T_0$. These issues will be further addressed
after the statements of the theorems.
\begin{thm}
(The increase of the binding energy)
Let Condition 1 be satisfied and $s_{disc}(h) \ne \emptyset$. Then
\begin{equation}
\Delta _E > e_0\ . \label{f.6.3}
\end{equation}
\end{thm}
\begin{thm}
(Enhanced binding)
Let Condition 1 be fulfilled, and let $V(|x|)= \beta V_0(|x|)$ be a
short-range potential with the properties described above.  Then
\begin{equation}
\beta _1< \beta _0\ . \label{f.7.1}
\end{equation}
\end{thm}

\noindent{\em Remark.} 
1. Let us first discuss theorems 1 and 2 for the spinless case $g=0$. 
The fact that $inf s(T_0)=E_0$ was proved by J. Fr\"ohlich for all $\alpha$. 
Recently, it was proved in [C] that for small $\alpha$, and
an ultraviolet cutoff $\chi\in C^1({\mathbb R}_+)$, 
$E_0$ is contained in the point 
spectrum of $T_0$. 
Thus, under these assumptions, the conditions of theorems 1 
and 2 are clearly fulfilled.

2. If $g=1$ (inclusion of particle spin), 
both parts of condition 1 remain
true. This follows from the generalizations of [C] outlined 
in the Appendix of the present work.

3. It may be useful to briefly comment on the case for models that
include an infrared regularization.
It follows from [C] that
Condition 1 is satisfied by models that are infrared regularized
by an infrared cutoff function in $A(0)$ that vanishes nowhere on $(0,\Lambda)$
(this technical requirement is used for the Ward-Takahashi identities
within the operator-theoretic renormalization group scheme of [C]).  
Consequently, theorems 1 and 2 hold for these models.
If the infrared cutoff vanishes in an open neighborhood of $\{0\}$,
or if it is incorporated by adding a small photon mass, 
the methods of [F] can be applied to verify
Condition 1. 

\begin{cor}
Let the ultraviolet cutoff $\chi (|k|)$ be fixed and have bounded first
derivatives. Then one can find a number $\alpha _0$  independent of the 
potential
$V(|x|)$, such that for all $0<\alpha <\alpha _0$ the following two statements
hold:
\begin{enumerate}
\item[i)]
if $e_0 \ne 0$ then $\Delta _E > e_0$;
\item[ii)]
if $V(|x|)= \beta V_0(|x|)$ is a short-range potential satisfying the same
condition as formulated above, then $\beta _1 <\beta _0$.
\end{enumerate}
\end{cor}

\section { Proof of Theorem 1. }
\setcounter{equation}{0}
To prove the theorem, we shall construct a trial function $\varphi \in {\mathcal
H}$, such that
\begin{equation}
(H\varphi ,\ \varphi )<(E_0 -e_0)\|\varphi \|^2 \ . \label{f.9.1.}
\end{equation}
Let $\psi _0$ be the ground state of the operator $T_0$
\begin{equation}
T_0\psi _0 = E_0\psi _0 \ , \qquad \|\psi _0\|_{{\mathcal H}_0}= 1 \ .
\label{f.9.2.}
\end{equation}
We will need the following fact.

\begin{lem}
Assume $\psi _0$ and $E_0$ as in Condition 1. Then, the relation
\begin{equation}
\|(P_f -\sqrt{\alpha}A(0))\psi _0\| \ne 0 
\label{f.6.2}
\end{equation}
holds \footnote{Notice that $(P_f
-\sqrt{\alpha}A(0))\psi _0$ is a 3-component vector with components \linebreak
$\psi
_{0i}\in {\mathcal H}_0$   $i=1, 2, 3.$ As usual $\|(P_f -\sqrt{\alpha}A(0))\psi
_0\|= \Big(\sum\limits_{i=1}^3 \| \psi _{0i}\|^2 _{{\mathcal H}_0}\Big)^{1/2}.$}.
\end{lem}

\noindent{{\em Proof.}}
By contradiction. Assume that (~\ref{f.6.2}) is incorrect. Then, 
\begin{equation}
P_f \psi_0= \sqrt\alpha A(0) \psi_0 \; ,
\label{contrassump}
\end{equation}
and the magnetic term yields
\begin{eqnarray*}
\sigma \cdot B_f(0) \psi_0 &=&
i\sigma \big(P_f\wedge A(0) + A(0)\wedge P_f\big) \psi_0 \\
&=&i\sigma \big(\frac{1}{\sqrt\alpha}P_f\wedge P_f + 
\sqrt\alpha A(0)\wedge A(0)\big) \psi_0  \\
&=&0 \; .
\end{eqnarray*}
Thus,
$$
T_0 \psi_0 = H_f \psi_0 = E_0 \psi_0 \; .
$$
But this yields a contradiction, since the only eigenvector of $H_f$ is 
the Fock vacuum, which 
fails to satisfy (~\ref{contrassump}) if $\alpha\neq0$. This establishes
the Lemma.
\\

Thus,  
$$
\Big(P_f - \sqrt{\alpha} A(0)\Big)\psi _0 = \sum\limits_{i=1}^3 e_i\psi
_{0i}\not\equiv 0\ ,
$$
where $e_i$  is the orthonormal basis in ${\Bbb R}^3$. Without loss of 
generality,
we assume $\psi _{01}\not\equiv 0$. Let $\tilde {\psi}_{01}$ be a fixed function 
in
$D(T_0)$, such that $\|\tilde {\psi}_{01}\|_{{\mathcal H}_0}= 1$ and
\begin{equation}
\Re (\tilde {\psi}_{01},\ \psi _{01})_{{\mathcal H}_0} \ge \frac1{2} \|
{\psi}_{01}\|_{{\mathcal H}_0}\ . \label{f.9.3}
\end{equation}
For our estimates it is more convenient to consider the functions $\psi _0$ and
$\tilde {\psi}_{01}$ in the coordinate (for photons) representation.
Let $\xi _i\in {\Bbb R}^3$ be the position vectors of photons. A function $\psi 
\in
{\mathcal H}_0$ can be written as
\begin{equation}
\psi =\  \mathrel{\mathop{\oplus}\limits_{n}}\psi _n(s,\ \xi _1,\ \ldots ,\ \xi
_n,\ \lambda _1,\ \ldots , \ \lambda _n)\ ,   \label{f.10.1}
\end{equation}
where $s$ is the electron spin and $\lambda _i$ are the polarization vectors of
photons.

For $x \in {\Bbb R}^3$ and an arbitrary element $\psi \in {\mathcal H}_0$
we define the operator $S_x$ and the function $\psi _x \in {\mathcal H}_0$ by 
the
formula
\begin{equation}
\psi _x=S_x\psi =\  \mathrel{\mathop{\oplus}\limits_{n}}\psi _n(s,\ \xi _1-x,\
\ldots ,\ \xi _n-x,\ \lambda _1,\ \ldots , \ \lambda _n)\ .   \label{f.10.2}
\end{equation}
Denote by $f_0(|x|)$ the real normalized eigenfunction of the operator $h$
corresponding to the lowest eigenvalue and let $f_1(x) \in C_0^2({\Bbb R}^3)$ be
the function with the following properties:
\begin{enumerate}
\item[i)]
$f_1(x)$ is real,
\item[ii)]
$\|f_1(x)\|=1$,
\item[iii)]
$f_1(x)$ is symmetric with respect to the reflections $x_2\leftrightarrow -x_2$ 
and
$x_3\leftrightarrow -x_3$ and antisymmetric with respect to the reflection
$x_1\leftrightarrow -x_1$,\footnote{Everywhere in the paper $x = (x_1,\ x_2,\
x_3)\in {\Bbb R}^3.$}
\item[iv)]
$$
\left( \frac{\partial f_0(|x|)}{\partial x_1},\ f_1(x)\right)> \frac1{2} \left\|
\frac{\partial f_0}{\partial x_1}\right\|\ .
$$
\end{enumerate}
Now we are ready to define the trial function $\varphi.$ Let $\eta$ be a real
valued parameter, which will be specified later and let
\begin{equation}
\varphi = f_0(x) \psi _{x0}+ i\eta f_1(x) \tilde {\psi}_{x01}\ , \label{f.11.1}
\end{equation}
where the functions $f_0$, $f_1$, $\psi _0$, $\tilde {\psi}_{01}$ are defined
above, $\psi _{x0}= S_x\psi _0$ and $\tilde {\psi}_{x01}= S_x\tilde 
{\psi}_{01}.$
Our next goal is to prove ~(\ref{f.9.1.})  for $\eta$ small and negative.

Obviously
\begin{eqnarray}
(H\varphi ,\ \varphi )& = & (Hf_0(x)\psi _{x0},\ f_0(x)\psi _{x0})-\eta
^2(Hf_1(x)\tilde {\psi}_{x01},\ f_1(x) \tilde {\psi}_{x01}) \nonumber\\
& & \\ \label{f.11.2}
& - &2\eta \Im (Hf_0(x)\psi _{x0},\ f_1(x)\tilde {\psi}_{x01})\nonumber
\end{eqnarray}
The second term in the right side of ~(\ref{f.11.2}) can be estimated from above 
as
$c_0\eta ^2$ with a constant $c_0$ independent of $\eta$. Let us evaluate the 
first
term. Notice that
\begin{equation}
pf_0(x)\psi _{x0}= \Big(pf_0(x)\Big)\psi _{x0}- f_0(x)P_f\psi _{x0}\ ,
\label{f.11.3}
\end{equation}
which implies
\begin{eqnarray}
(Hf_0(x)\psi _{x0}, \ f_0(x)\psi _{x0}) & = & \|pf_0(x) |^2 \| \psi
_{x0}\|^2_{{\mathcal H}_0}\\\label{f.11.4}
  +  (f_0(x)V(|x|),\ f_0(x))\| \psi _{x0}\|^2_{{\mathcal H}_0}
 & + & \|f_0(x)\|^2 (T\psi _0 ,\ \psi _0)_{{\mathcal H}_0} \nonumber\\
 & =  & ( E_0 - e_0)\|f_0(x)\psi _0\|^2 \nonumber
\end{eqnarray}
Here we also used the orthogonality
$$
\left( \frac{\partial f_0}{\partial x_i},\ f_0\right)= 0 \qquad i=1,\ 2,\ 3.
$$
To estimate the last term in ~(\ref{f.11.2}) recall that because of the symmetry
$$
(f_0,\ f_1)=0\ , \quad (V(|x|)f_0,\ f_1) =0,\ (\Delta f_0,\ f_1)=0\ ,\quad
\left( \frac{\partial f_0}{\partial x_i},\ f_1\right)=0\quad i=2,\ 3.
$$
Hence
\begin{eqnarray}
-2\eta \Im (Hf_0(x)\psi _{x0},\ f_1(x)\tilde {\psi}_{x01})&  = & 2\eta \left(
\frac{\partial f_0}{\partial x_1},\ f_1\right)\Re (\psi _{01},\ \tilde
{\psi}_{01})\nonumber\\
& \le & \frac{\eta}2 \|\psi _{01} \|_{{\mathcal H}_0}\left\|\frac{\partial
f_0}{\partial x_1} \right\|^2 \label{f.12.1}
\end{eqnarray}
for $\eta <0$.

Combining ~(\ref{f.12.1}), (3.7) and ~(\ref{f.11.3}) we arrive at
\begin{equation}
(H\varphi ,\ \varphi ) \le (E_0 -e_0)\|f_0 \|^2 +c_0\eta ^2 + \frac{\eta}2
\left\|\frac{\partial f_0}{\partial x_1} \right\|^2\| \psi _{01}\|_{{\mathcal
H}_0}\ . \label{f.12.2}
\end{equation}
To complete the proof of Theorem 1 notice that
$$
\|\varphi\|^2_{\mathcal H}= \|f_0(x) \|^2 + \eta ^2 \| f_1(x)\|^2 = \|f_0(x)\|^2
+\eta ^2
$$
and hence
\begin{equation}
(H\varphi ,\ \varphi ) \le (E_0 -e_0) \|\varphi  \|^2_{\mathcal H } +\eta 
^2[|E_0 -e_0| +c_0] +
\frac{\eta}6 \| \nabla f_0\|^2 \|\psi _{01}\|_{{\mathcal H }_0} \label{f.12.3}
\end{equation}
which for
$$
0>\eta >-\frac1{6} \|\nabla f_0(x)\|^2 \|\psi _{01}\|_{{\mathcal H}_0} [|E_0 
-e_0|
+c_0]^{-1}
$$
yields  ~(\ref{f.9.1.}).

\section{ Proof of Theorem 2 . }
\setcounter{equation}{0}
We shall prove that for $\beta =\beta _0$
\begin{equation}
\inf s(H) <E_0\ . \label{13.1}
\end{equation}
The statement of the theorem follows from ~(\ref{13.1}) and the variational
principle.

Let $0\le \gamma <1$, $V_0(|x|)$ satisfies the conditions of the theorem,
\begin{equation}
h_{\gamma , \beta _0} = -(1-\gamma )\Delta + \beta _0V_0(|x|)\ .  \label{13.2}
\end{equation}
The operator $h_{\gamma , \beta _0}$ does not have discrete eigenvalues for 
$\gamma
=0$, but for all $0<\gamma <1$ it has at least one real spherically symmetric
eigenfunction $f_{\gamma}(|x|)$.

Let $\varphi $ be the function defined by ~(\ref{f.11.1})  with $f_0(|x|)$ 
replaced
by $f_{\gamma}(|x|) $ and $f_1(|x|)$  replaced by $f_{\gamma ,1}(x) =
\frac{\partial}{\partial x_1}f_{\gamma}(|x|)$. We shall prove that for $\gamma 
>0$,
$|\eta |$ sufficiently small $(\eta <0)$, the inequality
\begin{equation}
(H\varphi ,\ \varphi )< E_0 \| \varphi \|^2 \  \label{13.3}
\end{equation}
is true.

To this end, first let us recall some properties of the functions $f_{\gamma}$ 
for
small $\gamma $ (see, for example, ~\cite{VZ}).

Let $\Bbb B$ be the closure of the space $C_0^{\infty}({\Bbb R}^3)$ in the norm
$\|\psi \|_{{\Bbb B}} = \|\nabla \psi  \|$. The equation
$$
-\Delta \psi + \beta _0 V_0(x) \psi = 0
$$
has a unique spherically symmetric solution ${\overline \psi} \in {\Bbb B}$,
$\|{\overline \psi } \|_{\Bbb B }=1$. This solution  (which is called the 
virtual
level or zero-resonance ) satisfies \nolinebreak : $\Delta {\overline \psi }$,
$V_0(|x|){\overline \psi }\in {\mathcal L}^2({\Bbb R}^3)$. Assume that the
eigenfunctions $f_{\gamma }(x) $  are normalized by the condition $\| \nabla
f_{\gamma} (x)\| =1$. Then ~\cite{VZ}
$$
\|\Delta f_{\gamma}- \Delta {\overline \psi} \| \to 0 \quad {\mbox as } \quad
\gamma \to 0\ ,
$$
which implies the inequality
\begin{equation}
\| \Delta f_{\gamma}\| \le 2\| \Delta {\overline \psi}\| \| \nabla f_{\gamma}\| 
=
C_0\| \nabla f_{\gamma} \| \label{14.1}
\end{equation}
for all $\gamma$ small and $C_0$ independent of $\gamma$.  Let us turn directly 
to
estimating the quadratic form $(H\varphi ,\ \varphi )$.

Similarly to ~(\ref{f.11.2}) we have
\begin{eqnarray}
(H\varphi , \ \varphi ) & = & (H f_{\gamma }(x)\psi _{x0}, \ f_{\gamma}(x) \psi
_{x0}) - \eta ^2 (H f_{\gamma ,1}(x)\tilde {\psi} _{x01}, \ f_{\gamma ,1}(x)
\tilde{\psi} _{x01})\nonumber\\
& - & 2\eta \Im (H f_{\gamma }(x)\psi _{x0}, \ f_{\gamma ,1}(x) \tilde {\psi}
_{x01})\ . \label{14.2}
\end{eqnarray}
For the second term in the right side of ~(\ref{14.2}) the bound holds :
\begin{eqnarray}
|\eta ^2 (H f_{\gamma ,1 }(x) \tilde {\psi} _{x01}, \ f_{\gamma ,1}(x) \tilde
{\psi} _{x01})|&  \le & \eta ^2 \{ C_1 \| f_{\gamma ,1}\|^2 + C_2 \|\nabla
f_{\gamma , 1} \|^2\}\nonumber\\
& \le & \eta ^2 [C_1 +2C_2C_0]\|\nabla f_{\gamma} \|^2 \ , \label{15.1}
\end{eqnarray}
where the constants $C_1$ and $C_2$ depend on $\tilde {\psi}_{01}$, but do not
depend on $f_{\gamma}$, and $C_0$ is the constant from ~(\ref{14.1}).

Analogously to ~(\ref{f.11.4}) and ~(\ref{13.2}) respectively
\begin{equation}
(H f_{\gamma }(x)\psi _{x0}, \ f_{\gamma}(x) \psi _{x0})=E_0 \|f_{\gamma}(x) 
\|^2 +
(h_{0, \beta _0}f_{\gamma}(x),\ f_{\gamma}(x)) \label{15.2}
\end{equation}
and
\begin{eqnarray}
-2\eta \Im (H f_{\gamma }(x)\psi _{x0}, \ f_{\gamma , 1}(x) \tilde {\psi} 
_{x01}) &
\le & -\frac{|\eta |}{2}\left\|\frac{\partial f_{\gamma}}{\partial 
x_1}\right\|^2
\|\psi _{01} \|_{{\mathcal H}_0} \nonumber\\
& = &  -\frac{|\eta |}{6} \|\psi _{01} \|_{{\mathcal H}_0}\|\nabla f_{\gamma} 
\|^2
\ , \label{15.3}
\end{eqnarray}
recalling that $\|\psi _0\|_{{\mathcal H}_0} = 1$.

Combining ~(\ref{14.2}) with ~(\ref{15.1})--(\ref{15.3}) we obtain
\begin{eqnarray}
(H\varphi ,\ \varphi ) & \le & E_0 \|f_{\gamma }(x) \|^2 + (h_{0, \beta
_0}f_{\gamma }(x) ,\ f_{\gamma}(x)) \nonumber\\
& + & \left\{\eta ^2 [C_1 +2C_2C_0] - \frac{|\eta |}6 \|\psi _{01}\|_{{\mathcal
H}_0}  \right\} \|\nabla f_{\gamma}(x)\|^2 \\ \label{15.4}
& \le & E_0\| \varphi \|^2 + (1-\delta )\|\nabla f_{\gamma} \|^2 + \beta
_0(V_0f_{\gamma}(x),\ f_{\gamma}(x))\nonumber \ ,
\end{eqnarray}
where
\begin{equation}
\delta =  \frac{|\eta |}6 \|\psi _{01}\|_{{\mathcal H}_0}-\eta ^2 
\Biggl[\frac1{3}
|E_0 |  + C_1 + 2C_2C_0 \Biggr]\ . \label{16.1}
\end{equation}
Here we used the relation
$$
\|\varphi \|^2_{\mathcal H} = \|f_{\gamma}(x) \|^2 + \eta ^2 \| f_{\gamma, 
1}\|^2
\|\tilde {\psi} _{01}\|_{{\mathcal H}_0}^2 =   \|f_{\gamma}(x) \|^2+ \frac1{3} 
\eta ^2
\|\nabla f_{\gamma}\|^2\ .
$$
Notice, that for $|\eta |$ small $\delta >0$, and to complete the proof of the
theorem it suffices to take $0<\gamma \le \delta $.

\section{Appendix}
\setcounter{equation}{0}

In this appendix, we comment on the following statements that were central for
the applications of theorems 1 and 2 in the case $g=1$.
For $\alpha$ sufficiently small:
\begin{itemize}
\item[(i)]
The global minimum of
inf spec$\{T_P\}$ for $P\in{\Bbb R}^3$ is attained at $P=0$.
\item[(ii)]
There exists a ground state $\psi_0\in{\mathcal H}_0$ of $T_0$   for  $g=1$.
\end{itemize}
Let us comment on the proof. We recall that
$$
        T_P = \big( P - P_f - \sqrt\alpha A\big)^2 + ig\sqrt\alpha \,
        \sigma\cdot\big(P_f \wedge A + A \wedge P_f\big) + H_f \; ,
$$
where $A\equiv A(0)$.
Let us to begin with include an artificial infrared regularization
in the quantized electromagnetic vector potential, which
acts like a momentum cutoff at small $\rho>0$ (some
requirements on its precise form are formulated in \cite{Ch}),
by which we substitute $A\rightarrow A(\rho)$ (under a slight
abuse of notation), and $T_P\rightarrow T_P(\rho)$.
Then, in addition to (i) and (ii), we claim that
\begin{itemize}
\item[(iii)]
For all $\rho>0$, and $|P|\geq0$ and $\alpha$ sufficiently small,
$$
        E(P,\rho) := {\rm inf\ spec} \{ T_P(\rho) \}  
$$
is an eigenvalue, whose eigenspace in ${\mathcal H}_P\cong{\Bbb 
C}^2\otimes{\mathcal F}$ has dimension $2$. Assume $\psi_P(\rho)\in{\Bbb 
C}^2\otimes{\mathcal F}$ is an eigenvector. Then, 
if $P=0$,
$\psi_0(\rho)$ tends to a ground state $\psi_0(0)\in{\Bbb C}^2\otimes{\mathcal 
F}$ in the limit $\rho\rightarrow0$. We note that the last statement is false
in the case $|P|>0$;
if $|P|>0$, $T_P(0)$ fails to have a ground state in ${\Bbb C}^2\otimes{\mathcal 
F}$.
\item[(iv)]
For some $\delta>0$, $0<P_c<1$ and $\alpha$ sufficiently small,
\begin{equation}
        \Big|\partial_P^b\Big(E(P,\rho)-P^2\Big)\Big| \leq C \alpha^\delta
        \label{secderest}
\end{equation}
uniformly for $\rho\geq0$, with $b=0,1,2$, and all $P$, $|P|<P_c$. 
\end{itemize}

The detailed proof of these results will be published separately.
The degeneracy of the ground state energy has recently been
proved by F. Hiroshima and H. Spohn, ~\cite{HS2}, for the case where the photons 
have a small mass.
We will here briefly sketch the strategy, which is
an extension of \cite{Ch}. It uses the operator-theoretic
renormalization group based on the smooth Feshbach map,
\cite{bcfs1,bcfs2,Ch}, and its framework can be roughly outlined as follows.
One introduces a certain Banach space ${\mathcal W}$ of bounded operators acting 
on the Hilbert space ${\bf 1}_2\otimes\chi(H_f<1){\mathcal H}_P$ (more 
precisely, one considers a particular Banach space of generalized Wick kernels 
that parametrizes such operators, but for simplicity, we do not make this 
distinction here), and makes a careful choice of a small polydisc ${\mathcal 
P}\subset{\mathcal W}$. Furthermore, one introduces a renormalization map 
${\mathcal R}:{\mathcal P}\rightarrow{\mathcal P}$, constructed by way of the 
isospectral smooth Feshbach map, and a rescaling transformation. The idea then 
is to focus on the dynamical system $({\mathcal P},{\mathcal R})$. A key 
property of ${\mathcal R}$ is that it is contractive on a subspace of ${\mathcal 
P}$ of finite codimension. Using the smooth Feshbach map, one associates 
$T_P(\rho)$ to an element  $H^{(0)}\in{\mathcal P}$, and considers the orbit 
$(H^{(n)})_{n\in{\Bbb N}_0}$ under ${\mathcal R}$ that emanates from this 
initial condition. The elements $H^{(n)}$ of this orbit are called {\em 
effective Hamiltonians}, where  $n$ is the {\em scale}, and in particular, they 
are mutually isospectral in the sense of the Feshbach theorem, \cite{bcfs1}.  
The fixed point of ${\mathcal R}$ on this orbit corresponds to the effective 
Hamiltonian in the scaling limit, $H^{(\infty)}$, and by isospectrality of the 
smooth Feshbach map, its spectral properties are directly related to those of 
$T_P(\rho)$.

The main ingredients in this construction are\footnote{T.C. thanks J. Fr\"ohlich 
for pointing out this key fact.}, in addition to the
arguments developed in \cite{Ch},  {\em parity invariance}, and {\em 
irrelevance} of the $B$-field
operator in renormalization group terminology. 
Indeed, under parity inversion, $x\rightarrow-x$, we have
$$
        P\rightarrow-P \; \; , \; \;
        P_f\rightarrow - P_f \; \; , \; \;
        A(\rho) \rightarrow -A(\rho) \; ,
$$
with respect to which $T_P(\rho)$ is evidently invariant.
The most general form of the effective Hamiltonian in the scaling limit is 
\begin{eqnarray*}
        H^{(\infty)} &=&  \alpha^{(\infty)}(P,\rho) H_f
        + \beta^{(\infty)}(P,\rho) P \cdot P_f \\
        &&+ \mu^{(\infty)}(P,\rho)
        \sigma \cdot P_f + 
        \nu^{(\infty)}(P,\rho) \sigma \cdot P \; ,
\end{eqnarray*}
where the coefficients $\alpha^{(\infty)}(P,\rho)$, $\beta^{(\infty)}(P,\rho)$, 
$\mu^{(\infty)}(P,\rho)$, and
$\nu^{(\infty)}(P,\rho)$ transform trivially under spatial rotations and 
reflections,
and are uniformly bounded in $\rho\geq0$.
Uniform boundedness with respect to $\rho\geq0$ is in part due to the 
irrelevance of the $B$-field operator,
and absence of interactions is due to the infrared regularization.

The renormalization map ${\mathcal R}$ is constructed in a manner that it
commutes with parity inversion, thus all $H^{(n)}$,  $n\in{\Bbb N}_0$, and 
$H^{(\infty)}$
are parity invariant. However, under parity reversal,
$\sigma\cdot P$ and $\sigma\cdot P_f$ change their signs. Therefore,
$\nu(P,\rho)=\mu(P,\rho)=0$, which implies that the ground state energy of
$H^{(\infty)}(P,\rho)$, of value $0$, is doubly degenerate.
Owing to the isospectrality properties of the smooth Feshbach map,
the same statement applies to $T_P(\rho)$. This proves (iii).

For the proof of  (i) and (iv), we remark that combining parity invariance with 
the generalized Ward-Takahashi identities of \cite{Ch}, it can be shown that the 
$\sigma^0$-component of the interaction  in $H^{(n)}$ is marginal, where 
$n\in{\Bbb N}_0$,
while the $\sigma^i$-components, for $i=1,2,3$, are irrelevant.
Hence, the study of marginal operators in \cite{Ch} can be straightforwardly 
adapted to the current problem,
and the corresponding results are valid even for $g=1$.
This immediately implies (i) and (iv). 

To prove (ii), we note that (iv) implies
$$
        \big|\partial_P E(P,\rho)\big|\geq (1-C \alpha^\delta) |P| \; ,
$$
which is bounded away from 0 for all $0\leq |P|\leq P_c$, for $C$
independent of $\rho\geq0$.
Thus, $E(P,\rho)$ has no minima away from 0 for
$0\leq |P|\leq P_c$.
Furthermore, writing $H_0:= H_f + (P-P_f)^2$ and $T_P=H_0+W$, we consider
$$
        T_P = H_0 + (H_0+\alpha)^{\frac{1}{2}}
        \Big((H_0+\alpha)^{-\frac{1}{2}}W(H_0+\alpha)^{-\frac{1}{2}}\Big)
        (H_0+\alpha)^{\frac{1}{2}} \; .
$$
From \cite{Ch,GLL} follows that
$\|(H_0+\alpha)^{-\frac{1}{2}}W(H_0+\alpha)^{-\frac{1}{2}}\|\leq C\sqrt\alpha$,
hence $T_P\geq C (H_0-\alpha)$ for $C\geq\frac{1}{2}$, and $\alpha$ sufficiently 
small.
Since ${\rm inf\ spec} \{H_0\}\geq C_1 P^2_c$ for $|P|\ge P_c$ with some 
$C_1>0$, it is evident that for all $\rho\geq0$,
$|E(P,\rho)|$ has its global minimum at $P=0$, such that (ii) follows.
\bigskip
\bigskip

\noindent {\it Acknowledgments:} S. Vugalter would like to thank H. ~Spohn for
attracting his attention to the results of ~\cite{F}. T. ~Chen is deeply
indebted to J. Fr\"ohlich for his advice and important comments. He also thanks 
H. ~Spohn for discussions. This work was partially supported by the European project HPRN-CT-2002-0027.

\bigskip

\end{document}